\documentclass[journal,twoside]{IEEEtran}
\usepackage{academicons}

\usepackage{cite}
\usepackage{graphicx}
\usepackage{comment}
\usepackage{balance}
\usepackage{acronym}
\usepackage{amsmath}
\usepackage{tabularx}
\usepackage{multirow}
\usepackage{amssymb}
\usepackage[usenames,dvipsnames]{xcolor}
\usepackage{caption,subcaption}
\usepackage{booktabs}
\usepackage{bbding}

\usepackage{url}
\usepackage[colorlinks=true]{hyperref}

\def\BibTeX{{\rm B\kern-.05em{\sc i\kern-.025em b}\kern-.08em
    T\kern-.1667em\lower.7ex\hbox{E}\kern-.125emX}}
    
\title{Audio-Visual Approach for Multimodal Concurrent Speaker Detection\thanks{The work was partially supported by a grant from the Audition Project, Data Science Program, Council of Higher Education, Israel, and by the European Union’s Horizon 2020 Research and Innovation Programme under Grant Agreement No. 871245.}
}

\author{Amit Eliav and Sharon Gannot,~\IEEEmembership{Fellow,~IEEE, 
\thanks{The authors are with Bar-Ilan University, Israel.  e-mail: \texttt{\{amiteli,sharon.gannot\}@biu.ac.il}.}
}
}

\acrodef{STFT}{short-time Fourier transform}
\acrodef{ISTFT}{inverse short-time Fourier transform}
\acrodef{TF}{Time-Frequency}
\acrodef{RTF}{Relative Transfer Function}

\acrodef{fps}{Frames Per Second}

\acrodef{AR}{Augmented-Reality}

\acrodef{LCMV}{Linearly Constrained Minimum Variance}

\acrodef{CSD}{Concurrent Speaker Detection}
\acrodef{VAD}{Voice Activity Detection}
\acrodef{OSD}{Overlapped Speech Detection}
\acrodef{SCD}{Speaker Change Detection}

\acrodef{NLP}{Natural Language Processing}

\acrodef{LSTM}{Long Short-Term Memory}
\acrodef{BLSTM}{Bidirectional Long Short-Term Memory}
\acrodef{RNN}{Recurrent Neural Networks}
\acrodef{CNN}{Convolutional Neural Networks}

\acrodef{GMM}{Gaussian mixture model}

\acrodef{ViT}{Vision Transformer}
\acrodef{CLS}{Class Token}
\acrodef{MHA}{multi-head attention}
\acrodef{TCN}{Temporal Convolutional Networks}
\acrodef{AST}{Audio Spectrogram Transformer}

\acrodef{CE}{Cross-Entropy}
\acrodef{CS}{Cost-Sensitive}
\acrodef{FL}{Focal-Loss}
\acrodef{LS}{Label-Smoothing}

\acrodef{mAP}{mean Average Precision}

\acrodef{SOTA}{state-of-the-art}
\acrodef{AST}{Audio Spectrogram Transformer}
\acrodef{ViT}{Vision Transformer}

\acrodef{LiDAR}{Light Detection and Ranging}

\begin{document}


\maketitle

\begin{abstract}
\label{abstract}
\acf{CSD}, the task of identifying active speakers and their overlaps in an audio signal, is essential for various audio applications, including meeting transcription, speaker diarization, and speech separation. This study presents a multimodal deep learning approach that integrates audio and visual information. The proposed model utilizes an early fusion strategy, combining audio and visual features through cross-modal attention mechanisms with a learnable [CLS] token to capture key audio-visual relationships.

The model is extensively evaluated on two real-world datasets, the established AMI dataset and the recently introduced EasyCom dataset. Experiments validate the effectiveness of the multimodal fusion strategy. An ablation study further supports the design choices and the model's training procedure.
As this is the first work reporting \ac{CSD} results on the challenging EasyCom dataset, the findings demonstrate the potential of the proposed multimodal approach for \ac{CSD} in real-world scenarios.
\end{abstract}

\section{Introduction}
\label{Introduction}
\acf{CSD} entails detecting active speakers and overlapping speech within an audio signal. \ac{CSD} classifies audio segments into three categories: 1) no speech activity (noise-only), 2) single-speaker activity, and 3) concurrent-speaker activity.
Accurate \ac{CSD} is crucial for various speech-processing applications, including audio scene analysis, meeting transcription, speaker counting and diarization, speech detection, and speech separation. 
A \ac{CSD} model is also advantageous for addressing ``cocktail party'' scenarios by analyzing signals from multiple microphones. A notable example is provided in \cite{8553564,Schwartz2024CSD}, where a multichannel \ac{CSD} model is incorporated into the design of an \ac{LCMV} beamformer. This model acts as a control mechanism to identify relevant time frames for estimating the fundamental components of the \ac{LCMV} beamformer, specifically its steering vectors and the spatial noise correlation function.

Two tasks closely related to \ac{CSD} are \ac{VAD} and \ac{OSD}. \ac{VAD} categorizes audio into active speech or non-active speech, while \ac{OSD} differentiates between overlapping and non-overlapping speakers. All three tasks are formally defined in Sec.~\ref{Problem Formulation}.
In studies \cite{8462548} and \cite{9053096}, the \ac{OSD} task was addressed using an \ac{LSTM} model. The work in \cite{cornell:hal-02908241} employs a \ac{TCN}-based model to tackle \ac{VAD}, \ac{OSD}, and a combined \ac{VAD}+\ac{OSD} task, which is equivalent to \ac{CSD}. Additionally, \cite{CORNELL2022101306} utilizes a Transformer-based model for these tasks, while \cite{zheng2021beamtransformer} applies a multichannel Transformer specifically for the \ac{OSD} task.
The recent work in \cite{lebourdais2023joint} addresses \ac{VAD}, \ac{OSD}, and the combined task using WavLM \cite{9814838} and \ac{TCN}.
In \cite{10094972}, a multi-task model is introduced for \ac{VAD}, \ac{OSD}, and \ac{SCD}, utilizing a fine-tuned `wav2vec 2.0' architecture \cite{NEURIPS2020_92d1e1eb}. Additionally, \cite{9414677} presents a model that combines speaker counting (up to two speakers), speech separation, and speech enhancement tasks. If a single speaker is detected, the model enhances that speaker; if overlapping speakers are detected, it first separates them before enhancing each one. Studies such as \cite{yousefi2021real} and \cite{kanda2020joint} employ attention mechanisms and \acp{CNN} jointly for tasks like speaker counting, speech recognition, and speaker identification in overlapped speech scenarios.
In our recent work \cite{eliav2024concurrent}, we presented an audio-only transformer-based \ac{CSD} model for both single- and multi-microphone audio data, presenting its effectiveness over 3 real-world datasets. This study also explores three different merging strategies for multi-microphone data. Building on these insights, we apply a similar merging methodology in this paper, as our focus remains on multi-microphone data.
`Pyannote' \cite{bredin2021end} is a Python library offering a variety of models for audio-related tasks, including speaker diarization, \ac{VAD}, and \ac{OSD}. It uniquely serves as the only publicly available package that allows for directly analyzing the datasets we investigate, thereby facilitating comparisons with our findings. For other comparisons, we rely on the results reported in the respective papers.

Despite these recent advances, the \ac{CSD} task remains challenging due to the inherent complexities involved in analyzing human speech. Variations in accent, pitch range, and speaking style across different individuals can make the accurate identification and detection of active speakers difficult. Additionally, real-world scenarios are often characterized by environmental noise and reverberation, further contributing to the difficulty of this problem. Consequently, \ac{CSD} continues to be an active area of research, with ongoing efforts aimed at developing more robust and accurate methods to handle this task and its related \ac{VAD} and \ac{OSD} tasks.

In this study, we introduce a deep learning approach for multimodal audio-visual models aimed at addressing the \ac{CSD} task.
Multimodal models have demonstrated improvements over single-modality models by integrating information from multiple sources, a process known as fusion \cite{PORIA201798}. These models are widely used in various applications, including audio-related tasks like audio-visual target speaker extraction \cite{10447462}, and vision tasks such as fusing \ac{LiDAR} and camera data \cite{10380551}. Combining both modalities can enhance a model's accuracy by providing a more comprehensive and robust representation of the environment.
While audio data may be affected by surrounding acoustic noise, video data tends to be more resilient, potentially capturing speakers even in noisy environments with minimal visual interference. However, relying solely on video data for a \ac{CSD} model is constrained by the camera's field of view, potentially missing speakers outside its scope.

Our research was motivated by our participation in the EU Horizon2020 project ``Socially Pertinent Robots in Gerontological Healthcare'' (SPRING)\footnote{\url{https://spring-h2020.eu/}}, aimed at developing assistive robots for healthcare applications, with other potential applications for public spaces like airports, malls, hospitals, or homes. The project involved multiple scientific disciplines and eight European partners. The audio pipeline of SPRING \cite{alameda2024socially} includes tasks such as speech detection, enhancement, speaker detection and localization, and speaker separation and extraction. \ac{CSD} is crucial in this pipeline, acting as a controller to determine which algorithm to activate for each segment. Additionally, the robot's multi-microphone array and cameras allow for the use of multi-modal (audio-visual) and multi-microphone processing to enhance audio-related tasks.

Multimodal datasets have become increasingly common, encouraging researchers to explore audio-visual approaches for the \ac{CSD} task. While many of the previously surveyed works relied solely on audio datasets, which can limit context capture, recent studies incorporate both audio and visual information for audio-related tasks. For instance, \cite{10064301} presents audio-visual and audio- and video-only models for the \ac{OSD} task. In \cite{mitchell2023study, 10447683}, an audio-visual model is introduced for speaker localization using the EasyCom dataset \cite{donley2021easycom}. Additional works, such as \cite{10184052, 10094836}, present additional audio-visual models for tasks like diarization, speech separation, dereverberation, and recognition.

Consequently, developing robust and accurate \ac{CSD} methods is critical to handling the inherent complexity and variability of real-world scenarios. By fusing information from both audio and visual modalities, we can potentially enhance the performance and robustness of \ac{CSD} models. This multimodal approach can provide complementary cues that address limitations present in individual modalities alone, leading to a more comprehensive understanding of the acoustic scene.

In this work, we propose an approach to address the \ac{CSD} task, introducing a deep-learning multimodal audio-visual model that effectively integrates multichannel audio with visual inputs.
We investigate both audio-only and visual-only models and compare them to the multimodal audio-visual scheme.
The model's architecture leverages an early fusion scheme, combining both modalities to enhance the classification capability.

Our main contributions are: 1) a novel multimodal model for the \ac{CSD} task leveraging state-of-the-art deep-learning models; 2) a comprehensive analysis of the proposed model with thorough comparisons to competing methods; 3) a training procedure utilizing different learning rates for the pre-trained backbone and other layers, along with audio and visual data augmentations, enhancing convergence and performance; 4) an evaluation of our model on two real-world datasets, including, to the best of our knowledge, the first reported \ac{VAD}, \ac{OSD}, and \ac{CSD} results for the recent EasyCom dataset \cite{donley2021easycom}.

The remainder of the paper is structured as follows.
Sec.~\ref{Problem Formulation} formulates the \ac{CSD} alongside the two related tasks of \ac{VAD} and \ac{OSD}.
Sec.~\ref{Proposed Model} presents our proposed model, including the audio and visual pre-processing, the data augmentation, the feature extraction backbones, and the fusion of the audio-visual data. Additionally, this section discusses the objective function and the loss regularization.
Sec.~\ref{Experimental study} covers the datasets employed in this work, the algorithm setup—including parameter choices and training procedures—and the model's performance. We thoroughly evaluated our model using various metrics and compared it with other available methods. Lastly, this section presents the ablation study conducted to examine the training process and two alternative model architectures.

\section{Problem Formulation}
\label{Problem Formulation}
Let $\mathbf{X}_A\in\mathbb{R}^{N\times \Tilde{L}}$ represent the audio data, where $N$ is the number of microphones, and $\Tilde{L}$ is the total data length in samples.
Let $\mathcal{X}_V\in\mathbb{R}^{\Tilde{F}\times C\times H\times W}$ represent the visual data, where $\Tilde{F}$ is the number of frames, $C$ is the number of channels (e.g., $C=3$ for RGB data), and $(H, W)$ is the image resolution.

Denote a single frame image as $\mathcal{X}_V^f\in\mathbb{R}^{C\times H\times W}$, with $f \in [1,\Tilde{F}]$. For each of these video frames, the corresponding audio frame-level data is denoted as $\mathbf{X}_A^f\in\mathbb{R}^{N\times T_f}$, where $T_f$ is the number of audio samples with a duration corresponding to a single video frame. 
Specifically, in the AMI dataset, the video frame rate is 20~fps, and the audio sampling rate is 16~kHz, yielding exactly $T_f = 800$ audio samples per video frame. In the EasyCom dataset, the video frame rate is 25~fps, and the audio was resampled to 16~kHz, resulting in $T_f = 640$ audio samples per video frame.

While our main focus is on the \ac{CSD} task, we begin by defining the two related and commonly addressed speaker detection tasks: \acf{VAD} and \acf{OSD}.

\ac{VAD} is a binary classification task that distinguishes between speech and non-speech regions in an audio signal. The task is performed at the resolution of each video frame, with the corresponding audio. Formally, for each video frame $f \in [1,\Tilde{F}]$, the \ac{VAD} classifies the audio-visual data as indicated below:
\begin{equation} 
    \label{eq:vad}
    \mathrm{VAD}(\mathbf{X}_A^f, \mathcal{X}_V^f) = 
    \begin{cases}
     \textrm{Class \#0} &  \textrm{Non-speech activity}  \\
     \textrm{Class \#1} &  \textrm{Speech activity}   \\ 
    \end{cases}.
\end{equation}
A time frame $f$ is marked as active if either a single speaker or multiple speakers are present.

\ac{OSD} is a similarly binary classification task that distinguishes between overlapping and non-overlapping speakers. Similar to \ac{VAD}, it is performed at the resolution of each video frame.
Formally, for each video frame $f \in [1,\Tilde{F}]$, \ac{OSD} classifies the audio-visual data as indicated below:
\begin{equation} 
    \label{eq: osd}
    \mathrm{OSD}(\mathbf{X}_A^f, \mathcal{X}_V^f) = 
    \begin{cases}
     \textrm{Class \#0} &  \textrm{Non-overlapped speech}  \\
     \textrm{Class \#1} &  \textrm{Overlapped speech}   \\
    \end{cases},
\end{equation}
where non-overlapping segments designate either noise-only or a single active speaker.

While \ac{VAD} and \ac{OSD} are fundamental to many audio processing systems, they have limitations in distinguishing between different signal types within the same class. In the case of \ac{VAD}, both single-speaker and overlapping-speaker speech are grouped as active speech despite their differing statistical behaviors. Similarly, though they represent distinct acoustic scenarios, \ac{OSD} treats noise-only and single-speaker activity as one class. By separating these cases into individual classes, \ac{CSD} enables finer-grained categorization, thereby enhancing the understanding and analysis of the acoustic scene.

The multimodal \ac{CSD} algorithm combines both the \ac{VAD} and \ac{OSD} tasks into a single multi-class classification task.
In the \ac{CSD} classification task, each video frame and its corresponding audio data (either single-microphone or multi-microphone) is classified into one of the three classes as indicated below, for $f \in [1,\Tilde{F}]$:
\begin{equation} 
    \label{eq: csd}
    \mathrm{CSD}(\mathbf{X}_A^f, \mathcal{X}_V^f) = 
    \begin{cases}
     \textrm{Class \#0} & \textrm{Noise only}\\
     \textrm{Class \#1} & \textrm{Single-speaker activity}\\ 
     \textrm{Class \#2} & \textrm{Concurrent-speaker activity}
    \end{cases}\,.
\end{equation}
Identifying and analyzing audio data in the context of the \ac{CSD} task presents significant challenges due to the inherent variability in speech and acoustics.
The distribution of statistical features within audio data can vary significantly based on the underlying acoustic scene. For example, Class \#0 (`Noise-Only') may include different noise types, each with unique statistical characteristics. Similarly, Class \#1 (`Single-speaker activity') faces challenges due to the diversity of human speech, as individual speakers have distinct accents, styles, and vocal traits that complicate accurate identification. Furthermore, Class \#2 (`Concurrent-speaker activity') adds complexity due to varying numbers of active speakers, resulting in a broader range of statistical properties.

In this work, we choose to split the input data into short segments, with each segment comprising 7 frames of video along with their corresponding audio data.
Each segment undergoes preprocessing to crop and extract only the faces, resizing them to a fixed size of $224\times224$ pixels, as detailed in Sec.~\ref{Pre_Processing}.
Each 7-frame clip of cropped faces is considered a stream. Consequently, the number of visual streams in each segment depends on the number of detected faces in the given clip.
Thus, the visual input to our model is of shape $\#\textrm{Streams} \times 7 \times 3 \times 224 \times 224$.
The shape of the audio input to our model is $N \times L$, where $L$ is the length, in samples, corresponding to 7 frames of video, which may vary with the video frame rate.
Finally, our model receives the audio-visual input and outputs 7 labels corresponding to the 7 input video frames, classifying each frame into one of the three \ac{CSD} classes.

\section{Proposed Model}
\label{Proposed Model}
\begin{figure}[t]
    \centering
      \includegraphics[width=0.43\textwidth]{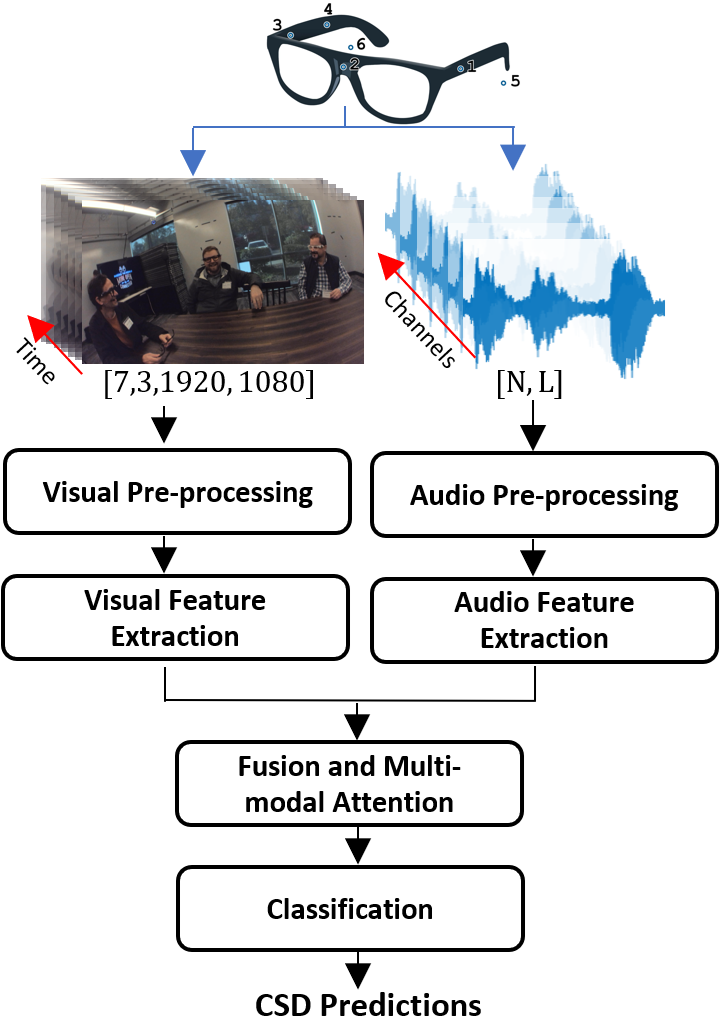}
      \setlength{\belowcaptionskip}{0pt}
      \caption{Overview of the proposed model, including input data, pre-processing, feature extraction, fusion, and classification. Illustrating the pipeline from raw data to \ac{CSD} predictions, demonstrated for the EasyCom dataset.}
      \label{fig:model_overview}
\end{figure}

The proposed model comprises several components, including feature extraction backbones, audio and visual processing blocks, and a fusion scheme. We utilize pre-trained audio and video models as backbone feature extractors.
An overview of the proposed model is depicted in Fig.~\ref{fig:model_overview}. It illustrates the pipeline from raw data to \ac{CSD} predictions, demonstrated for the EasyCom dataset.
%
The audio backbone extracts features from the input multichannel audio data using a pre-trained HuBERT model \cite{hsu2021hubert}.
The visual backbone extracts features from the visual data, using streams of cropped faces derived from the original video, as detailed in Sec.~\ref{Pre_Processing}. A pre-trained R3D-18 model \cite{Tran_2018_CVPR} serves as the backbone feature extractor for each video stream.

Additionally, we consider a fusion technique to combine the audio and visual modalities. We explore both early and late fusion approaches, along with other mechanisms like \ac{MHA}, to facilitate information transfer between modalities. Ultimately, the model employs early fusion techniques to jointly process the data and perform the \ac{CSD} classification task.

\subsection{Pre-Processing and Input data}
\label{Pre_Processing}
Both the audio and visual data undergo distinct pre-processing pipelines. 

The microphone signals are first resampled to 16 kHz to align with the audio backbone's sampling rate. For the visual data, a stream is extracted for each detected face using a YOLOv8 model \cite{Jocher_YOLO_by_Ultralytics_2023} trained for face detection.\footnote{The model's weights are available on \url{https://github.com/akanametov/yolo-face}, we used the `yolov8n-face.pt' model.} Each stream is resized to a resolution of $224 \times 224$. The maximum number of streams depends on the dataset; for the EasyCom dataset, it is 8, and for the AMI dataset, it is 7. If a segment has fewer detected streams than the maximum, it is zero-padded. For the AMI dataset, all 4 ``Closeup'' cameras are utilized, concatenating their detected streams.

The output labels are derived from the transcribed datasets, with a resolution of a single video frame: 0.04 seconds for the EasyCom dataset (25~fps) and 0.05 seconds for the AMI dataset (20~fps). We use 7 video frames along with the corresponding audio data as input to the model. Consequently, the dimensions of the inputs are $N \times L$ for the audio tensor and $\#\textrm{Streams} \times 7 \times 3 \times 224 \times 224$ for the visual tensor, where $L = 5600$  for EasyCom and $L = 4480$ for AMI.
The output prediction is a tensor of size $7 \times 3$, representing the probabilities for the three classes corresponding to each of the seven input video frames.

\subsection{Data Augmentation and Balancing}
\label{data_augmentation}
Most available datasets for the \ac{CSD} task exhibit significant class imbalance, reflecting typical patterns in natural human conversations, as illustrated in Table~\ref{tab: Class frequency}. This imbalance is addressed during training through various techniques, including tuning the loss function, as discussed in Sec.~\ref{Objectives}, and employing data augmentation methods.
%
%
\begin{table}[htbp]
    \caption{Class frequency [\%] in the training set for all datasets. The number of frames is given in million [M].\\ 
    Dataset$^\dagger$ for a balanced and augmented dataset.}
    \label{tab: Class frequency}
    \centering
    \begin{tabular}{p{1.8cm}cccc}
         \toprule
         Dataset/Class& \#0 [\%] & \#1 [\%] &  \#2 [\%] & \#Frames [M]\\
         \midrule
         AMI          & 16.8    & 71.8   & 11.4   & 7.1 \\
         AMI$^\dagger$      & 40.3    & 29.4   & 30.3   & 7.8 \\
         \midrule
         EasyCom      & 30.5    & 58.2   & 11.3   & 0.255\\
         EasyCom$^\dagger$   & 22      & 39     & 39     & 1.2 \\
         \bottomrule
    \end{tabular}
\end{table}
Data augmentation and balancing are crucial in classification tasks to prevent the model from favoring the majority class. Augmentation serves as an effective strategy for both audio and visual data, enhancing the diversity of the training set and improving model robustness.

To create a more balanced dataset, the training set was adjusted to achieve a more uniform class distribution. The process began by including all segments containing class \#2 (“Concurrent-speaker activity”) frames. Additional frames were then randomly sampled from classes \#0 (“Noise only”) and \#1 (“Single-speaker activity”).

The datasets summarized in Table~\ref{tab: Class frequency} include two variants of both the AMI and EasyCom datasets. The first variant is the original dataset, which reflects the natural class distribution and is heavily imbalanced toward class \#1 (“Single-speaker activity”). The second variant, marked as Dataset\textsuperscript{\dag}-with $\textrm{Dataset}\in\{\textrm{AMI},\textrm{EasyCom}\}$-is derived from the original data. It consists of several balanced sub-datasets, each generated as described above, followed by an augmentation process. This approach increases the diversity of the resulting balanced and augmented training set.

For the audio data, we apply two augmentation procedures: 1) pitch shifting in the time domain and 2) spectral masking in the frequency domain, which can mask the entire time frame (full-band) or use time-frequency patches.

\begin{figure}[ht]
    \centering
    \begin{subfigure}[t]{0.23\textwidth}
        \centering
        \includegraphics[height=2.5cm]{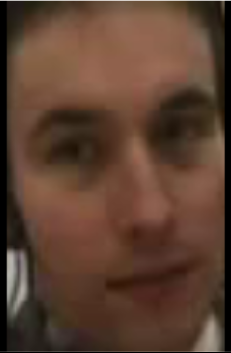}
        \caption{Original Frame: Cropped face from the original video frame using the YOLOv8 model trained with face detection.}
    \end{subfigure}
    \hfill
    \begin{subfigure}[t]{0.23\textwidth}
        \centering
        \includegraphics[height=2.5cm]{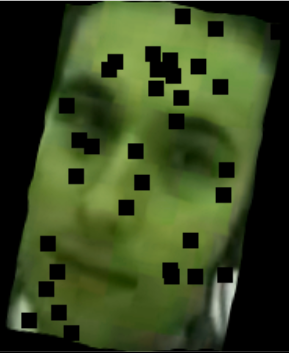}
        \caption{
        Frame augmentation: Combination of a horizontal flip, color jitter, and masking patches.
        \vspace{-0.25cm}
        }
    \end{subfigure}
    \hfill
    \begin{subfigure}[t]{0.23\textwidth}
        \centering
        \includegraphics[height=2.5cm]{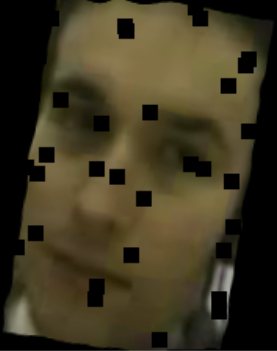}
        \caption{Frame augmentation: Combination of a horizontal flip, Gaussian blur, and masking patches.}
    \end{subfigure}
    \hfill
    \begin{subfigure}[t]{0.23\textwidth}
        \centering
        \includegraphics[height=2.5cm]{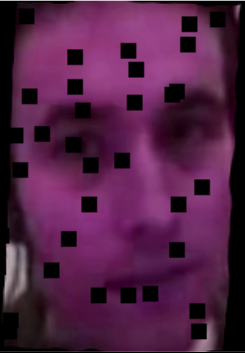}
        \caption{Frame augmentation: Combination of randomly adjusted sharpness, color jitter, and masking patches.}
    \end{subfigure}
    \caption{Visual data augmentations: An example of a frame from the AMI dataset alongside its various augmentations.}
    \label{fig:visual_augmentations}
\end{figure}

For the visual data, we utilize several augmentation techniques, including `Random Rotation,' `Elastic Transform,' `Random Horizontal Flip,' `Color Jitter,' `Grayscale,' `Gaussian Blur,' and `Random Adjust Sharpness.' Additionally, we implement random masking by setting patches of pixels to zero. Specifically, around 45 patches of size $10\times10$ pixels are randomly distributed and masked across each video frame.
%
%
Figure~\ref{fig:visual_augmentations} depicts examples of visual data augmentations.

\subsection{Architecture - Backbones, Audio- and Visual-Blocks}
\label{architecture_audio_and_visual_block}
The audio backbone is based on a pre-trained HuBERT model \cite{hsu2021hubert}, which is used as a feature extractor from each of the microphone input data. The audio backbone receives the preprocessed tensor of shape $(N \times L)$, and the audio backbone is applied to each microphone signal. The last Transformer layer of the HuBERT model is used to extract the tokens. There are $S'$ tokens of dimension $768$ extracted from each audio channel. The extracted tokens from the multichannel data are concatenated along the first dimension, resulting in a $(S\times 768)$ features tensor, where $S=N\cdot S'$.
Concatenation along the microphone dimension is backed by our recent study, which compares three merging strategies for multichannel audio data concatenation for the \ac{CSD} task \cite{eliav2024concurrent}.

The visual backbone processes the cropped face streams after preprocessing, as detailed in Sec.~\ref{Pre_Processing}. It utilizes a pre-trained R3D-18 model \cite{Tran_2018_CVPR} as a feature extractor for each stream. Each stream generates a feature vector of dimension $512$, and the extracted features are concatenated along the stream dimension, resulting in a tensor of size $(\#\textrm{Streams} \times 512)$.

\begin{figure}[htp]
    \centering
      \includegraphics[width=0.36\textwidth]{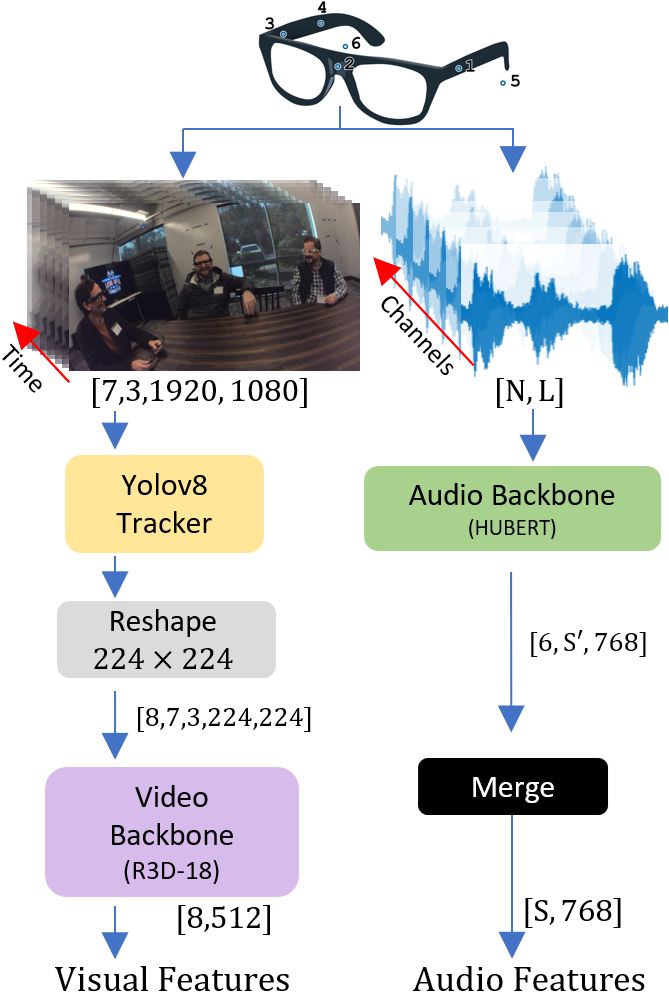}
      \setlength{\belowcaptionskip}{0pt}
      \caption{Audio-Visual feature extraction demonstrated for the EasyCom dataset.
      $S=N \cdot S\,'$, where $S\,'$ is the number of extracted tokens from the audio segments, and $N=6$ for the EasyCom dataset.}
      \label{audio_video_feature_extraction}
   \end{figure}

These initial steps of preprocessing and feature extraction from each modality are presented in Fig.~\ref{audio_video_feature_extraction} and demonstrated for the EasyCom dataset. The two backbones are used to extract the feature vectors of the two modalities, of shapes $(S\times 768)$ and $(\#\textrm{Streams}\times 512)$, for the audio and visual modalities, respectively.

The audio and visual blocks, as shown in Fig.~\ref{fig: fusion_classification}, share a similar architecture, consisting of normalization layers, \ac{MHA}, and fully connected layers.
The attention mechanism, which was first used in the context of \ac{NLP} \cite{9222960, vaswani2017attention} was proven to be beneficial for audio-related tasks, e.g., for the \ac{AST} model \cite{gong21b_interspeech} in audio classification applications.
These audio and visual blocks are used to enhance the features of their respective modalities and to contribute to the fusion scheme, as outlined in Sec.~\ref{architecture_fusion}.
%
%

\subsection{Architecture - Fusion and Classification}
\label{architecture_fusion}
\begin{figure*}[htp]
    \centering
      \includegraphics[width=0.99\textwidth]{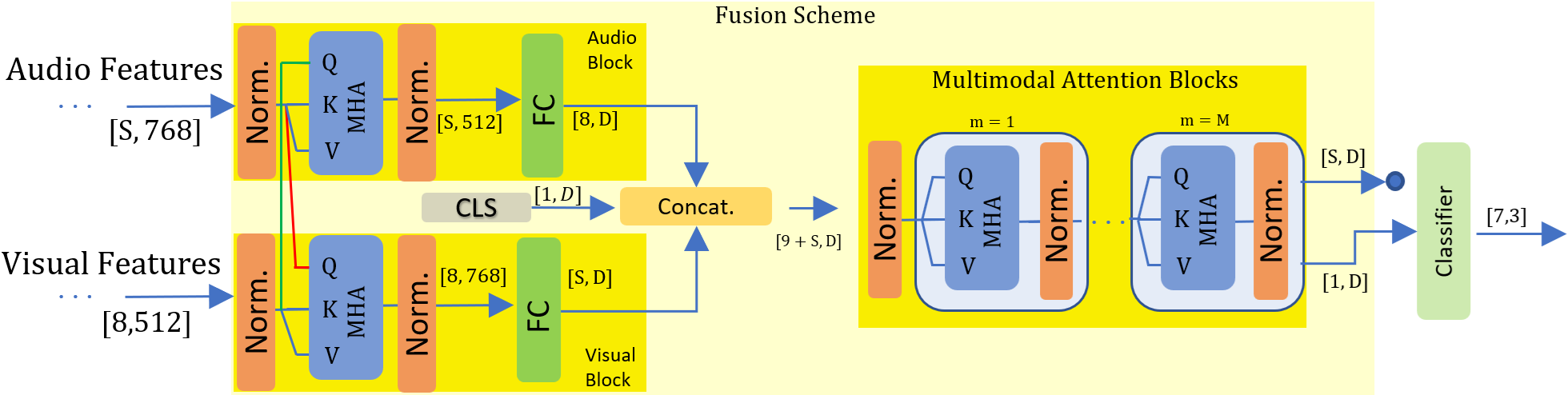}
      \setlength{\belowcaptionskip}{0pt}
      \caption{The audio-visual fusion scheme, the multimodal \ac{MHA} blocks, and the classification layer demonstrated for the EasyCom dataset.}
      \label{fig: fusion_classification}
   \end{figure*}

Effectively combining the audio and visual modalities is essential for achieving an accurate classification in the \ac{CSD} task.
The fusion process allows the model to leverage the information from both audio and video inputs, enhancing its ability to distinguish between the three \ac{CSD} classes.
This section details the architecture design of the fusion process and the subsequent classification of audio-visual data. We now discuss each component used for fusion and classification in detail. The audio-visual fusion scheme, the multimodal \ac{MHA} blocks and the classification layer are presented in Fig.~\ref{fig: fusion_classification}.
%
%

\noindent\textbf{Normalization Layers}:
The first step in fusing the audio-visual modalities occurs in the audio and visual blocks, where a normalization layer is applied to each modality's tokens. Normalization layers are employed separately for each modality, both before and after the \ac{MHA} layer, to ensure that the extracted tokens are on a similar scale. This mitigates the potential impact of differing value ranges across modalities on the subsequent layers.

\noindent\textbf{Fully Connected Layers - A Common Embedding Space}:
Each feature extraction backbone produces tokens in different dimensions—768 for audio and 512 for visual data. Fully-connected layers are used for each modality to project the tokens into a common dimension $D$, ensuring that the tokens from different modalities are represented in a shared embedding space.

\noindent\textbf{\ac{MHA} Configuration}:
The \ac{MHA} mechanism is defined by several key parameters, with the most relevant to our choice of the fusion architecture being the input tensors Query, Key, and Value, $Q, K, V$, respectively.
In the context of our fusion design, we need to determine which tokens from each modality will be used as the $Q$, $K$, and $V$ input tensors for the \ac{MHA}. Specifically, the $Q$ input can come from either the same modality or the other modality's tokens.

\noindent\textbf{Early Fusion and Concatenation with [CLS] Token}:
The \ac{MHA} is used with a cross-modality strategy, where each modality uses the other modality's tokens as the $Q$ input tensor.
The \ac{MHA} layer passes and extracts the information within each modality's tokens as well as across the two modalities, thereby initiating the early fusion of the audio and visual data.
The projected tokens from the two modalities are then concatenated with a class token [CLS] (of the same dimension), which is an additional learnable token. The concatenated tokens are fed into $M$ multimodal attention blocks, each consisting of a \ac{MHA} mechanism and normalization layers. Each block captures cross-modal interactions among the fused tokens, followed by a normalization layer to stabilize the process. These stacked blocks refine the cross-modal representations, allowing the model to capture relationships and dependencies between the two modalities.

\noindent\textbf{Classification Layer}:
The classifier uses only the token corresponding to the [CLS] token as input, producing a tensor of size $(7 \times 3)$ for predicting the 7 output label probabilities for each class. The [CLS] token mechanism is designed to ensure the classification process is unbiased toward any specific input tokens, as discussed in \cite{dosovitskiy2020image} in the context of Transformer models. This approach was also proven effective in our recent audio-only \ac{CSD} study \cite{eliav2024concurrent}.

Additionally, we evaluated three alternative fusion strategies: early fusion without the [CLS] token and late fusion approaches with and without the [CLS] token.
These alternatives are further discussed in the ablation study in Sec.~\ref{ablation_study}, which provides additional support for our chosen fusion scheme.

\subsection{Objective Functions}
\label{Objectives}
Since the model is designed for the \ac{CSD} task, the common choice for the loss function is the \ac{CE} loss. To address the classification imbalance among the three classes, class weights\footnote{\texttt{https://towardsdatascience.com/class-weights-\\for-categorical-loss-1a4c79818c2d}} are incorporated into the loss calculation, assigning higher weights to the underrepresented classes.

Additionally, \acf{LS} \cite{muller2019does} is applied to the ground-truth labels, which introduces a small degree of noise and prevents the model from overconfident predictions. \ac{LS} has been shown to improve generalization performance and mitigate overfitting, as was also used in our recent work \cite{eliav2024concurrent} and was proven beneficial.

By combining \ac{CE} loss with class weighting and \ac{LS}, the training objective aims to optimize the model's ability to accurately classify the data across both modalities while accounting for samples that are less accurately classified and promoting better generalization.

Besides the combination of \ac{CE} loss, class weighting, and \ac{LS}, which we consider as the baseline loss formulation, we explored alternative loss functions and regularizations to train our model and address the class imbalance issue. Specifically, we explored two additional losses as regularizers to the baseline loss, namely \ac{CS} loss \cite{galdran2020cost} and focal loss \cite{Lin_2017_ICCV}.
The \ac{CS} loss is designed to penalize different types of errors during model training and has proven beneficial in our recent work \cite{eliav2024concurrent}.
The Focal-Loss is an extension of the known \ac{CE} loss designed to address class imbalance by focusing on hard-to-classify examples.
Incorporating the \ac{CS} loss resulted in a less stable training process. Additionally, the focal loss did not exhibit a clear impact on the model's performance, failing to provide substantial improvements over the baseline loss formulation.
As a result, we opted for the combination of \ac{CE} loss, class weighting, and \ac{LS}, which proved to be the most effective approach for optimizing the audio-visual \ac{CSD} model.

\section{Experimental study}
\label{Experimental study}
In this section, we describe the experimental study carried out to validate the performance of the proposed algorithm.
\subsection{Datasets}
We evaluated the performance of our model using two real-world datasets, the EasyCom dataset \cite{donley2021easycom} and the AMI dataset \cite{10.1007/11677482_3}.
Both datasets use a microphone array, EasyCom with 6 microphones and AMI with 8. However, they differ in the available cameras: EasyCom uses a single wide-angle camera, while AMI uses multiple cameras, including room overview and close-up cameras.

The AMI \cite{10.1007/11677482_3} dataset comprises 100 hours of meeting recordings featuring English speakers (both female and male). Participants were recorded in three different room environments with various acoustic setups. The dataset includes an 8-microphone array and several cameras, including a close-up camera for each participant, a corner camera, and an overview camera. For this work, all sessions utilized the four close-up cameras, as detailed in Section~\ref{Pre_Processing}.

The EasyCom dataset \cite{donley2021easycom} is a relatively new dataset recorded using Meta's \ac{AR} glasses, which feature a 6-microphone array and a wide-angle single camera. Collected in a noisy environment, imitating a restaurant, the dataset includes multiple English speakers engaging in conversations during various tasks.
Two key challenges arise from the use of the \ac{AR} glasses worn by one participant during the meetings. First, the audio amplitude of the wearer's speech is significantly higher than that of other active participants due to the proximity of the microphone array. Second, rapid head movements by the wearer lead to fast changes in the visual data, causing shifts in the perceived locations of the speakers relative to the glasses' viewpoint, which also affects the acoustic characteristics of the speakers' voices. These simultaneous movements of both the speakers and the recording device contribute to the complexity of this multimodal dataset.
Since the EasyCom dataset is limited in volume, with only about 6 hours of data and highly unbalanced classes, we utilized multiple instances of the training set with various augmentations, as described in Sec.~\ref{data_augmentation}. The dataset was split into segments (7-frame-long clips) with a substantial overlap of 6 frames to enhance training diversity and mitigate class imbalance.

Both datasets exhibit a significant class imbalance favoring classes \#0 and \#1 (`Noise only' and `Single-speaker activity'). This imbalance reflects the natural dynamics of human conversation, where participants usually take turns speaking, resulting in minimal overlapping speech among multiple individuals.
This imbalance must be addressed during model training. We used three methods:
First, we applied data augmentation, as described in Sec.~\ref{data_augmentation}.
Second, creating training sets with a more balanced representation among the classes, as described in Sec.~\ref{data_augmentation}.
Third, we tuned the loss function, as outlined in Sec.~\ref{Objectives}.
The distribution of the different classes is depicted in Table~\ref{tab: Class frequency} for both the original datasets and for the datasets after balancing and augmentation.

\subsection{Algorithm Setup}
\label{Setup}
We used the architecture described in Section~\ref{Proposed Model} and shown in Fig.~\ref{audio_video_feature_extraction} and Fig.~\ref{fig: fusion_classification}, with the early fusion scheme and the [CLS] token mechanism. 
The fusion dimension is set to $D=512$, and the number of multimodal attention blocks is set to $M=4$.
To account for the varying number of detected video streams per segment, we padded all segments to a fixed number of streams (as described in Section~\ref{Pre_Processing}).
Additionally, to address the order of the detected faces, we randomly shuffled the streams within each segment during training. This approach ensures that the model does not become biased towards the order of the detected streams or the zero-padded streams.

In the model training, we used the Adam optimizer with a different learning rate for the different layers of the model, a weight decay of $1e^{-9}$, and a batch size of 64.
The learning rate was set to $1e^{-7}$ for the audio backbone, $1e^{-6}$ for the visual backbone, and $1e^{-4}$ for the rest of the layers (the audio and visual blocks, the fusion scheme and the classification layer).
This differential learning rate facilitates fine-tuning of the large pre-trained backbones at a slower pace, preventing drastic alterations to the learned representations while allowing the fusion and classification components to adapt more quickly to the target \ac{CSD} task.

Initially, an attempt was made to freeze the audio and visual backbones without retraining them, but this resulted in poor overall performance (as shown in Table~\ref{tab: Ablation_study}). This may be attributed to the backbones not being specifically trained for the \ac{CSD} task, resulting in suboptimal feature representations for the fusion and classification stages, as well as the downstream task.

To mitigate overfitting due to the model's substantial number of parameters—94 million for the audio backbone, 33 million for the visual backbone, and 8 million for the remaining layers, totaling approximately 135 million parameters—we limited the training process to a modest number of epochs, typically between 3 and 5. The exact number of epochs depended on the specific dataset under consideration.

\subsection{Competing Methods}
\label{sec:competing}
We compare our results with several leading methods, including audio-only, visual-only, and audio-visual models.
In \cite{zheng2021beamtransformer}, a multichannel audio-only Transformer model is used for the task of \ac{OSD}.
Similarly, \cite{CORNELL2022101306} presents a multichannel audio-only Transformer model for the tasks of \ac{OSD}.
In our recent work \cite{eliav2024concurrent}, we applied an audio-only transformer-based model to tackle the \ac{CSD} task using both single- and multi-microphone measurements. That method was originally evaluated on the AMI dataset. In this contribution, we use \cite{eliav2024concurrent} as a baseline after re-training it with the EasyCom dataset.

We compare our results with the visual-only and audio-visual models for the task of \ac{OSD} reported in \cite{10064301}, both of which use only single-microphone input from the AMI microphone array.
Another recent work, \cite{lebourdais2023joint}, addresses the \ac{VAD} and \ac{OSD} task by using WavLM \cite{9814838} and \ac{TCN}, with both single- and multi-channel audio-only variants. Notably, this work uses close-talk microphones, resulting in different acoustic conditions than the distant microphone array setup.
Finally, in \cite{10094972}, a fine-tuned `wav2vec 2.0' is employed for the tasks of \ac{VAD} and \ac{OSD} using audio-only data.
All these works were only applied to the AMI dataset. In all reported results in our comparative study, we relied exclusively on the results reported in the respective papers.

The publically available `Pyannote' Python toolkit \cite{bredin2021end}\footnote{Available on https://huggingface.co/pyannote} offers various speech-related models, including \ac{VAD} and \ac{OSD}. In our comparative study, we used the results as reported in \cite{bredin2021end} for the AMI dataset. 
The EasyCom dataset is relatively new, and to the best of our knowledge, no previous \ac{VAD}, \ac{OSD}, or \ac{CSD} results using this dataset have been reported in the literature.
We therefore used the `Pyannote' code to obtain the \ac{VAD} and \ac{OSD}. These classification results were then combined to synthetically generate the results for the \ac{CSD} task, as explained in the sequel.

Specifically, the \ac{VAD} model classifies audio into two categories: `0' for noise and `1' for speech activity (single or multiple speakers). Similarly, the \ac{OSD} model assigns `0' to noise or single-speaker activity and `1' to multiple active speakers. By summing the predictions from both models, we can synthesize the possible \ac{CSD} cases, as illustrated below:
\begin{equation} 
    \label{eq: csd_by_vad_osd}
    \mathrm{CSD}(\mathrm{VAD}, \mathrm{OSD}) = 
    \begin{cases}
     \textrm{$0_{\mathrm{VAD}} + 0_{\mathrm{OSD}}$} & \textrm{$0_{\mathrm{CSD}}$}\\
     \textrm{$0_{\mathrm{VAD}} + 1_{\mathrm{OSD}}$} & \textrm{No such case}\\
     \textrm{$1_{\mathrm{VAD}} + 0_{\mathrm{OSD}}$} & \textrm{$1_{\mathrm{CSD}}$}\\
     \textrm{$1_{\mathrm{VAD}} + 1_{\mathrm{OSD}}$} & \textrm{$2_{\mathrm{CSD}}$}
    \end{cases}.
\end{equation}
Additionally, we verified across the entire EasyCom dataset that the case where \ac{VAD} predicts `0' (indicating noise) and \ac{OSD} predicts `1' (indicating multiple active speakers) does not occur. This is a desirable outcome, as it ensures that the models consistently do not detect multiple speakers without speech activity.

These synthetically generated \ac{CSD} predictions enable us to compare our results for the EasyCom dataset across all three important tasks - \ac{VAD}, \ac{OSD}, and \ac{CSD}. In addition, we retrained our previous proposed model from \cite{eliav2024concurrent} using the EasyCom dataset and compared its performance to the proposed models in this paper.

\subsection{Results}
\label{Results}
Common metrics such as accuracy, precision, recall, F1-score, and \ac{mAP} are typically used to evaluate the performance of classification models. Additionally, a confusion matrix provides a detailed comparison between the ground-truth labels and the model's predicted labels, normalized as percentages relative to the ground-truth labels. These metrics enable a comprehensive assessment of our model's performance and facilitate comparisons with other methods, as the same metrics are reported in the respective articles.

Recall that the proposed model processes seven video frames along with their corresponding audio and generates output predictions for each frame. We noticed that the performance metrics are highest for the center frame (the fourth frame), making it the most reliable for classification. Consequently, this work reports results solely for the center frame, while the other six frames provide contextual information to classify the activity state more effectively.
During inference, the model still processes seven input frames and outputs predictions for all seven frames, but only the center prediction should be considered. The input window then slides by one frame to generate the prediction for the next center frame.

Table~\ref{tab: EasyCom model configurations comparison} presents the results for various model variants evaluated on the EasyCom dataset. We compare different configurations, including early and late fusion schemes and the integration of the [CLS] token. This comparative analysis aims to highlight the impact of the fusion strategy and the contribution of the [CLS] token on audio-visual \ac{CSD}.
Additionally, we compare the audio-visual variants with two audio-only models and a visual-only variant. The first audio-only model is derived from our recent work \cite{eliav2024concurrent} and has been retrained on the new EasyCom dataset. The second audio-only variant employs the architecture of our current proposed model but without the visual branch. Similarly, the visual-only variant is based on the proposed model, excluding the audio branch.
%
\begin{table*}[htbp]
    \centering
    \caption{A comparison of the proposed audio-visual model across four configurations, evaluating the performance on the \ac{VAD}, \ac{OSD}, and \ac{CSD} tasks. Accuracy (A), Precision (P), Recall (R), F1-score (F1), and \ac{mAP} (\%) measures are reported for the EasyCom dataset. 
    \textbf{Bold}: best overall, \underline{underlined}: best within modality.}
    \label{tab: EasyCom model configurations comparison}
\resizebox{\textwidth}{!}{
\begin{tabular}{llccccccccccccccc}\toprule
& \multicolumn{1}{c}{} & \multicolumn{5}{c}{\ac{VAD}} & \multicolumn{5}{c}{\ac{OSD}} & \multicolumn{5}{c}{\ac{CSD}} \\
\cmidrule(lr){3-7}\cmidrule(lr){8-12}\cmidrule(lr){13-17}
Modalities & Method & A & P  & R & F1 & \ac{mAP}   & A & P  & R & F1 & \ac{mAP}  & A & P  & R & F1 & \ac{mAP} \\
\midrule
\multirow{2}{*}{Audio} 
    & \cite{eliav2024concurrent}  & 74.1 & 73.5  & 74.1  & 72.5  & 87.5   & 81.6  & \underline{85.9}  & 81.6  & 83.5  & \underline{25.0}     & 59.5 & 62.9 & 59.5 & 60.2 & 66.3 \\
    & Audio-Block                 & \underline{76.8} & \underline{77.2}  & \underline{76.8}  & \underline{77.0}  & \underline{89.1}   & \underline{82.5}  & 85.5  & \underline{82.5}  & \underline{83.9}  & \underline{25.0}   & \underline{59.8} & \underline{64.9} & \underline{59.8} & \underline{61.0} & \underline{66.9} \\
\midrule
\multirow{1}{*}{Visual} 
    & Visual-Block               & 64.7 & 66.1  & 64.7  & 65.2  & 79.7   & 83.9  & 84.7  & 83.9  & 84.3  & 19.3   & 53.1 & 54.4 & 53.1 & 53.5 & 55.9 \\
\midrule
\multirow{4}{*}{Audio-Visual}
    & Early, w/o [CLS]               & 74.8 & 75.4  & 74.8  & 75.0  & 88.0   & 87.7  & 86.1  & 87.7  & \textbf{\underline{86.8}}  & 27.6   & 64.1 & 64.3 & 64.1 & 64.0 & 68.5 \\
    & \textbf {Early, with [CLS]}    & \textbf{\underline{79.0}} & \textbf{\underline{81.2}}  & \textbf{\underline{79.0}}  & \textbf{\underline{79.4}}  & \textbf{\underline{92.8}}   & \textbf{\underline{90.0}}  & 87.0  & \textbf{\underline{90.0}}  & 86.6  & \textbf{\underline{32.8}}   & \textbf{\underline{70.4}} & \textbf{\underline{69.6}} & \textbf{\underline{70.4}} & \textbf{\underline{67.9}} & \textbf{\underline{71.7}} \\
    & Late, w/o [CLS]             & 41.1          & 52.3  & 41.1  & 38.6  & 63.5   & 89.8  & 85.8  & 89.8  & 85.1  & 10.8   & 35.1 & 52.9 & 35.1 & 18.4 & 40.9 \\
    & Late, with [CLS]            & 77.5          & 78.4  & 77.5  & 77.7  & 90.4   & 82.6  & \textbf{\underline{87.4}}  & 82.6  & 84.4  & 31.3   & 61.5 & 67.7 & 61.5 & 62.5 & 71.0 \\
\bottomrule
\end{tabular}}
\end{table*}

Table~\ref{tab: EasyCom model configurations comparison} provides a comprehensive comparison of our proposed model across all three tasks. The results clearly demonstrate that the early fusion variant with the [CLS] token mechanism outperforms both the audio-only and video-only models, as well as the method presented in \cite{eliav2024concurrent}.

Table~\ref{tab: confusion matrices} presents the confusion matrices for both datasets, reporting the results of the best audio-visual model variant, which employs early fusion and the [CLS] token.
%
\begin{table}[htbp]
    \centering
    \caption{\ac{CSD} results: confusion matrices normalized to the ground-truth labels [\%]. `T' denotes true labels, while `P' indicates predicted labels.}
    \label{tab: confusion matrices}
\begin{tabular}{lccccccccc}\toprule
& \multicolumn{3}{c}{AMI} & \multicolumn{3}{c}{EasyCom}
\\\cmidrule(lr){2-4}\cmidrule(lr){5-7}
           T \textbackslash P & 0  & 1 & 2   & 0  & 1 & 2
           \\\midrule
0        & 89 & 8  & 3    & 81 & 15 & 4  \\
1        & 14 & 73 & 13   & 26 & 60 & 14 \\
2        & 3  & 38 & 59   & 16 & 42 & 42
\\\bottomrule
\end{tabular}
\end{table}
%
As shown in this table, the model performs well on class \#0 (`Noise only'), achieving high accuracy. However, for the more challenging class \#2 (`Concurrent-speaker activity'), the model accuracy (normalized to the true class) is only 42\% for the EasyCom dataset and 59\% for the AMI dataset.

A comparison of our best model variant with available methods, in terms of Accuracy, Precision, Recall, F1-score, and \ac{mAP}, is presented in Table~\ref{tab: comparison_AMI} and Table~\ref{tab: comparison_EasyCom} for the AMI and EasyCom datasets, respectively.

For the AMI dataset, we can directly compare our results with state-of-the-art methods since several previous studies have reported on the relevant metrics. However, most of these works focused on the \ac{OSD} task, so we adapted our multi-class \ac{CSD} classification results into a binary \ac{OSD} classification. This was achieved by aggregating the probabilities of classes \#0 and \#1.
For the EasyCom dataset, we used the `Pyannote' toolkit to extract predictions for all three tasks, as described in Sec.~\ref{sec:competing}. We also followed the same procedure of aggregating the relevant probabilities to obtain the \ac{VAD} and \ac{OSD} predictions from our \ac{CSD} model.

\begin{table}[htbp]
    \caption{A comparison between the proposed model and several competing methods in evaluating the performance on the \ac{OSD} task, including Accuracy (A), Precision (P), Recall (R), F1-score (F1) and \ac{mAP} in (\%) measures on the AMI dataset.
    \textbf{Bold}: best overall, \underline{underlined}: best within modality.}
    \label{tab: comparison_AMI}
    \centering
\resizebox{\columnwidth}{!}{
\begin{tabular}{p{1.2cm}cccccc}
 \toprule
 Modalities & Method & A & P & R & F1 & \ac{mAP} \\
  \midrule
      \multirow{8}{0pt}{Audio}
                    & \cite{zheng2021beamtransformer}             & N/A   & 87.8  & 87    & N/A   & N/A  \\
                    & \cite{CORNELL2022101306}                    & N/A   & 87.8  & 87    & N/A   & 60.3 \\
                    & \cite{eliav2024concurrent}                  & N/A   & \textbf{\underline{92.4}}  & 89    & N/A   & \textbf{\underline{73.1}} \\
                    & \cite{bredin2021end}                        & N/A   & 80.7  & 70.5  & 75.3  & N/A  \\
                    & \cite{10064301} (Single-Channel)            & N/A   & N/A   & N/A   & N/A   & 62.7 \\
                    & \cite{lebourdais2023joint} (close-talk mic) & N/A   & N/A   & N/A   & 80.4  & N/A  \\
                    & \cite{10094972}                             & \textbf{\underline{94.16}} & 79.04 & 79.38 & 79.21 & N/A  \\
                    & \textbf{Our Audio-Block}                    & 89.6  & 89.6  & \textbf{\underline{89.6}}  & \textbf{\underline{89.6}}  & 63   \\
  \midrule
      \multirow{2}{0pt}{Visual}
                    & \cite{10064301}             & N/A   & N/A   & N/A   & N/A   & 20   \\
                    & \textbf{Our Visual-Blcok}   & \underline{80.9}  & \underline{87.6}  & \underline{80.9}  & \underline{83.2}  & \underline{51.6} \\
  \midrule
      \multirow{2}{0pt}{Audio-Visual}
                    & \cite{10064301}           & N/A   & N/A   & N/A   & N/A   & \underline{67.2} \\
                    & \textbf{Our Audio-Visual} & \underline{85.4} & \underline{87.5} & \underline{85.4} & \underline{86.3} & 53.1 \\
 \bottomrule
\end{tabular}}
\end{table}


\begin{table*}[t]
    \centering
    \caption{A comparison between the proposed model and two available methods in evaluating the performance on the \ac{VAD}, \ac{OSD}, and \ac{CSD} tasks, including Accuracy (A), Precision (P), Recall (R), F1-score (F1) and \ac{mAP} in (\%) measures on the EasyCom dataset.}
    \label{tab: comparison_EasyCom}
\begin{tabular}{lccccccccccccccc}\toprule
& \multicolumn{5}{c}{\ac{VAD}} & \multicolumn{5}{c}{\ac{OSD}} & \multicolumn{5}{c}{\ac{CSD}}
\\\cmidrule(lr){2-6}\cmidrule(lr){7-11}\cmidrule(lr){12-16}
\parbox[c]{1.4cm}{Method}
           & A & P  & R & F1 & \ac{mAP}   & A & P  & R & F1 & \ac{mAP}  & A & P  & R & F1 & \ac{mAP}
           \\\midrule
\cite{eliav2024concurrent} (Audio-only) & 74.1  & 73.5  & 74.1  & 72.5  & 87.5   & 81.6  & 85.9  & 81.6  & 83.5  & 25     & 59.5 & 62.9 & 59.5 & 60.2 & 66.3 \\ 
\cite{bredin2021end} (Adapted, Audio-only) & 77.0  & 76.8  & 77.0  & 75.6  & N/A    & 88.8  & 86.1  & 88.8  & \textbf{87.0}    & N/A    & 66.9 & 66.8 & 66.9 & 64.8 & N/A  \\
Our Audio-Block             & 76.8  & 77.2  & 76.8  & 77.0  & 89.1   & 82.5  & 85.5  & 82.5  & 83.9  & 25.0   & 59.8 & 64.9 & 59.8 & 61.0 & 66.9 \\
Our Audio-Visual             & \textbf{79.0}  & \textbf{81.2}  & \textbf{79.0}  & \textbf{79.4}  & \textbf{92.8}   & \textbf{90.0}  & \textbf{98.0}  & \textbf{90.0}  & 86.6  & \textbf{32.8}   & \textbf{70.4} & \textbf{69.6} & \textbf{70.4} & \textbf{67.9} & \textbf{71.7} \\ 
\bottomrule
\end{tabular}
\end{table*}

When evaluating the AMI dataset, we found that the audio-visual model does not outperform other models, as shown in Table~\ref{tab: comparison_AMI}. Moreover, when comparing the audio-visual model to the audio-only model, incorporating visual information does not enhance performance and may even slightly degrade it.
In contrast, when applied to the EasyCom dataset, the audio-visual model exhibits clear improvements, surpassing both audio-only models in most metrics across all three tasks. This indicates that integrating audio and visual modalities is more effective in the challenging environments characteristic of the EasyCom dataset.

To gain deeper insight into the performance of the proposed model, we present a confusion matrix in Table~\ref{tab: EasyCom_CM_comparison}, comparing our best audio-visual model with the classification results from \cite{bredin2021end}.
%
%
\begin{table}[htbp]
    \centering
    \caption{EasyCom \ac{CSD} comparison: confusion matrix comparison between the available method \cite{bredin2021end} and our audio-visual (AV) model, as [\%] normalized to the ground-truth labels. `T'-true labels, `P'-predicted labels.}
    \label{tab: EasyCom_CM_comparison}
\begin{tabular}{lccccccccc}\toprule
& \multicolumn{3}{c}{Our AV model} & \multicolumn{3}{c}{\cite{bredin2021end}}
\\\cmidrule(lr){2-4}\cmidrule(lr){5-7}
           T \textbackslash P & 0  & 1 & 2   & 0  & 1 & 2
           \\\midrule
0        & 81 & 15 & 4    & 50 & 48 & 2  \\
1        & 15 & 60 & 14   & 10 & 87 & 3  \\
2        & 16 & 42 & 42   & 3  & 78 & 19
\\\bottomrule
\end{tabular}
\end{table}
%
%
Both Table~\ref{tab: comparison_EasyCom} and Table~\ref{tab: EasyCom_CM_comparison} illustrate the challenges posed by the EasyCom dataset, resulting in lower performance compared to the AMI dataset. However, our audio-visual model handles EasyCom more effectively, achieving higher values across most metrics. The confusion matrix reveals that the classification performance of \cite{bredin2021end} is heavily biased toward class \#1 (“Single-speaker activity”), whereas our model maintains a more balanced performance across all three classes.

\subsection{Ablation Study}
\label{ablation_study}
We conducted an ablation study to evaluate the impact of three key components on our proposed model's performance: one related to the training process and two concerning the model architecture. For the training process, as detailed in Sec.~\ref{data_augmentation}, we applied various data augmentation techniques to the training data and trained the model both with and without these augmentations to assess their effect on classification performance.

Regarding the model architecture, we investigated the effects of training versus freezing the backbone feature extraction models and the impact of different fusion strategies. Specifically, we examined two scenarios for backbone training: allowing the pre-trained backbone models to update during training with a different learning rate than the other layers, as discussed in Sec.~\ref{Setup}, and keeping the backbone weights fixed while only training the remaining model layers.

Table~\ref{tab: Ablation_study} presents the four combinations of data augmentation and backbone training evaluated on the EasyCom dataset.
%
%
\begin{table*}[h!!]
    \centering
    \caption{Ablation study: A comparison of the proposed audio-visual model with and without data augmentation and backbone training, evaluated using \ac{VAD}, \ac{OSD}, and \ac{CSD} task. We report on the following measures: Accuracy (A), Precision (P), Recall (R), F1-score (F1), and \ac{mAP} (\%) on the EasyCom dataset.}
    \label{tab: Ablation_study}
\resizebox{\textwidth}{!}{
\begin{tabular}{ccccccccccccccccc}\toprule
  \multicolumn{2}{c}{} & \multicolumn{5}{c}{\ac{VAD}} & \multicolumn{5}{c}{\ac{OSD}} & \multicolumn{5}{c}{\ac{CSD}}
\\\cmidrule(lr){3-7}\cmidrule(lr){8-12}\cmidrule(lr){13-17}
           \parbox[c]{1.4cm}{\centering Data \\ augmentations} & \parbox[c]{1.4cm}{\centering Backbone \\ training}  & A & P & R & F1 & mAP   & A & P & R & F1 & mAP  & A & P & R & F1 & mAP
           \\\midrule
\XSolidBrush &  \XSolidBrush   & 64.9 & 42.1 & 64.9 & 51.1 & 71.5    & 88.1 & 81.1 & 88.1 & 84.6 & 22.2   & 59.0 & 60.0 & 59.0 & 60.0 & 68.5  \\
\checkmark   &  \XSolidBrush   & 77.9 & 79.2 & 77.9 & 78.3 & 91.7    & 86.5 & 86.9 & 86.5 & \textbf{86.7} & 32.3   & 65.6 & 68.0 & 65.6 & 65.8 & 71.5  \\
\XSolidBrush &  \checkmark     & 77.5 & 79.3 & 77.5 & 77.9 & 91.2    & 83.5 & 86.5 & 83.5 & 84.8 & 29.2   & 64.1 & 67.2 & 64.1 & 64.6 & \textbf{71.7}  \\
\checkmark   &  \checkmark     & \textbf{79.0} & \textbf{81.2} & \textbf{79.0} & \textbf{79.4} & \textbf{92.8}    & \textbf{90.0} & \textbf{87.0} & \textbf{90.0} & 86.6 & \textbf{32.8}   & \textbf{70.4} & \textbf{69.6} & \textbf{70.4} & \textbf{67.9} & \textbf{71.7}    
\\\bottomrule
\end{tabular}}
\end{table*}
Applying data augmentation and training the backbone networks clearly enhances overall performance. However, when the backbones were trained at the same learning rate as the rest of the model, rapid overfitting occurred, causing the model to consistently predict a single class. Consequently, we have opted not to include these results in the experimental study.

\begin{figure*}[h!]
     \centering
     \begin{subfigure}[t]{0.99\textwidth}
         \centering
\includegraphics[width=0.97\textwidth]{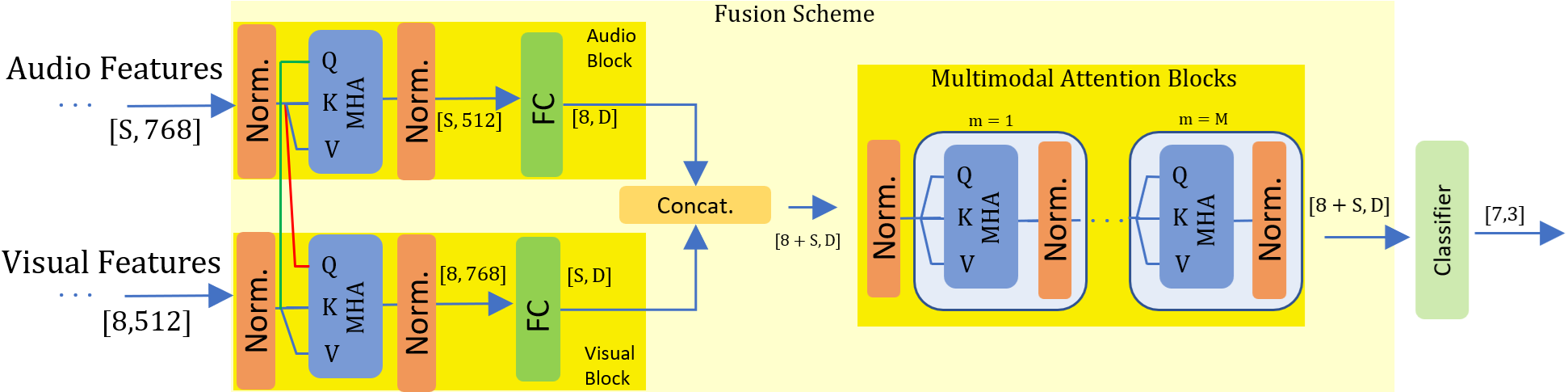}
      \caption{Early fusion scheme without [CLS] token mechanism.}
      \label{fig:architecture_fusion_v1}
     \end{subfigure}
     \hfill
     \begin{subfigure}[t]{0.99\textwidth}
         \centering
\includegraphics[width=0.97\textwidth]{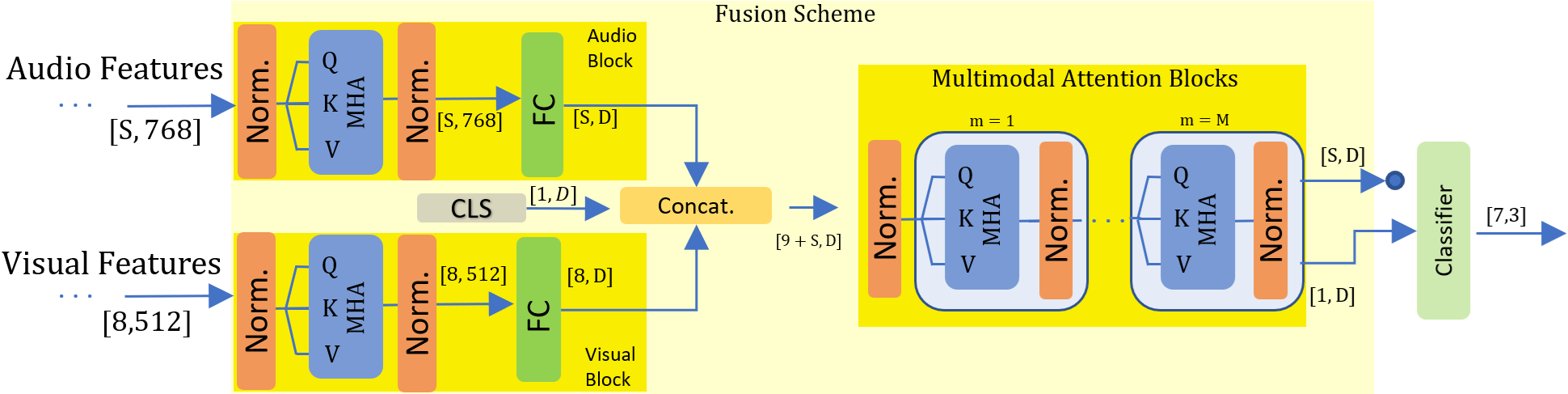}
      \caption{Late fusion scheme with [CLS] token mechanism.}
      \label{fig:architecture_fusion_v4}
     \end{subfigure}
     \hfill
     \begin{subfigure}[t]{0.99\textwidth}
         \centering
\includegraphics[width=0.97\textwidth]{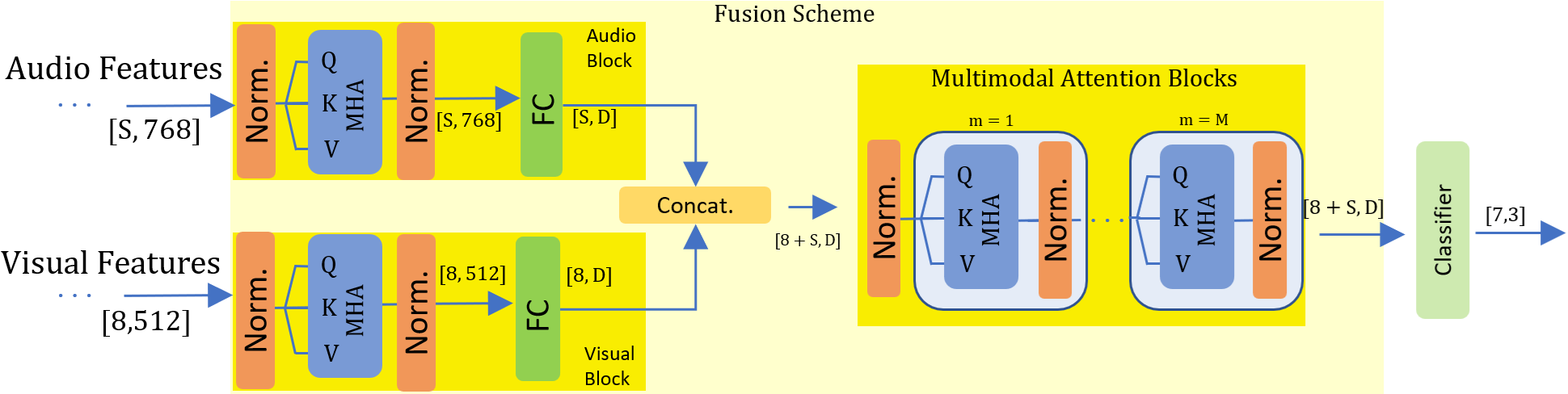}
      \caption{Late fusion scheme without [CLS] token mechanism.}
      \label{fig:architecture_fusion_v3}
      \end{subfigure}
      
      \caption{Three alternative fusion schemes demonstrated for the EasyCom dataset.}
\end{figure*}

Our proposed model employs an early fusion scheme in conjunction with the \ac{CLS} token mechanism. To support this architectural choice, we evaluated the effect of the \ac{CLS} token as well as different fusion strategies on the model's performance.
Specifically, we considered three configurations: early fusion without the \ac{CLS} token (Fig.~\ref{fig:architecture_fusion_v1}), late fusion with the \ac{CLS} token (Fig.~\ref{fig:architecture_fusion_v4}), and late fusion without the \ac{CLS} token (Fig.~\ref{fig:architecture_fusion_v3}).

In the late fusion variants—both with and without the [CLS] token—the overall fusion scheme and architecture closely resemble those of the proposed early fusion model. The primary distinction lies in the configuration of the Multi-Head Attention (MHA) layers at the beginning of the fusion process. Typically, MHA layers process three inputs: query ($Q$), key ($K$), and value ($V$) tensors. In the late fusion approach, each modality branch uses its own feature vector for all three tensors ($Q$, $K$, and $V$). In contrast, the early fusion variants implement a cross-modality input strategy, where each modality's MHA receives feature vectors from the other modality as the $Q$ input tensor. This cross-modality configuration, also employed in \cite{10064301} for an OSD model, facilitates the early integration of audio and visual modalities, enabling the model to more effectively capture cross-modal relationships and dependencies at the feature level.

Excluding the [CLS] token from the fusion scheme caused the classifier to receive an excessively large feature vector, resulting in an overly complex fully connected classification layer. Consequently, this approach was deemed less desirable. Additionally, late fusion strategies ultimately underperformed compared to the early fusion approach. These factors led us to adopt the early fusion scheme incorporating the [CLS] token mechanism for our proposed model. A detailed analysis is presented in Sec.~\ref{Results} and Table~\ref{tab: EasyCom model configurations comparison}.
%
%
%

\section{Conclusions}
\label{Conclusions}
In this study, we introduce a comprehensive deep learning approach to the \ac{CSD} task by leveraging multimodal audio-visual models. Our research contributes to the Socially Pertinent Robots in Gerontological Healthcare (SPRING) project, aiming to enhance the robustness and accuracy of \ac{CSD} in complex, real-world environments, including public spaces and interactive meeting settings.

We evaluated our proposed models on two real-world datasets, AMI and EasyCom, encompassing various audio-visual scenarios. Utilizing the YOLO model for video preprocessing, we extracted face streams to improve the accuracy of visual feature extraction. Additionally, we employed state-of-the-art audio and video backbone architectures to ensure effective feature representation from both modalities. The model architecture integrates these features through a carefully designed fusion strategy, enabling seamless integration and leveraging information from both audio and visual inputs.

Our model adopts an early fusion strategy, combining audio and visual features through cross-modal attention mechanisms and refining the joint representations via stacked multimodal attention blocks. By incorporating the [CLS] token, the model effectively captures the audio-visual relationships pertinent to the \ac{CSD} task.

Results indicate that our multimodal approach achieved slightly inferior performance on the AMI dataset compared to competing methods. However, it demonstrated significant improvements on the more challenging EasyCom dataset, highlighting the effectiveness of our approach in complex environments.

Ablation studies confirmed the critical role of data augmentation techniques and the use of differential learning rates for the audio and visual backbones compared to the other layers. These strategies substantially enhanced the model's performance, providing valuable insights into optimizations for both the training process and model architecture.

As multimodal technologies evolve and audio-visual data become increasingly abundant, our study demonstrates the significant potential of fusing audio and visual information. This offers an innovative method for audio-visual \ac{CSD} in increasingly complex acoustic environments. Additionally, we present the first reported results on the challenging EasyCom dataset for the three critical tasks of \ac{VAD}, \ac{OSD}, and \ac{CSD}, providing valuable insights into the performance of our approach in real-world scenarios.
%
%

 
\bibliographystyle{IEEEtran}
\bibliography{my_library}

\begin{thebibliography}{10}
\providecommand{\url}[1]{#1}
\def\UrlFont{\rmfamily}
\providecommand{\newblock}{\relax}
\providecommand{\bibinfo}[2]{#2}
\providecommand\BIBentrySTDinterwordspacing{\spaceskip=0pt\relax}
\providecommand\BIBentryALTinterwordstretchfactor{4}
\providecommand\BIBentryALTinterwordspacing{\spaceskip=\fontdimen2\font plus
\BIBentryALTinterwordstretchfactor\fontdimen3\font minus \fontdimen4\font\relax}
\providecommand\BIBforeignlanguage[2]{{%
\expandafter\ifx\csname l@#1\endcsname\relax
\typeout{** WARNING: IEEEtran.bst: No hyphenation pattern has been}%
\typeout{** loaded for the language `#1'. Using the pattern for}%
\typeout{** the default language instead.}%
\else
\language=\csname l@#1\endcsname
\fi
#2}}

\bibitem{8553564}
S.~E. Chazan, J.~Goldberger, and S.~Gannot, ``{LCMV} beamformer with {DNN}-based multichannel concurrent speakers detector,'' in \emph{26th European Signal Processing Conference (EUSIPCO)}, 2018, pp. 1562--1566.

\bibitem{Schwartz2024CSD}
\BIBentryALTinterwordspacing
A.~Schwartz, O.~Schwartz, S.~E. Chazan, and S.~Gannot, ``Multi-microphone simultaneous speakers detection and localization of multi-sources for separation and noise reduction,'' \emph{EURASIP Journal on Audio, Speech and Music}, vol.~50, Oct. 2024. [Online]. Available: \url{https://doi.org/10.1186/s13636-024-00365-3}
\BIBentrySTDinterwordspacing

\bibitem{8462548}
N.~Sajjan, S.~Ganesh, N.~Sharma, S.~Ganapathy, and N.~Ryant, ``Leveraging {LSTM} models for overlap detection in multi-party meetings,'' in \emph{IEEE International Conference on Acoustics, Speech and Signal Processing (ICASSP)}, 2018, pp. 5249--5253.

\bibitem{9053096}
L.~Bullock, H.~Bredin, and L.~P. Garcia-Perera, ``Overlap-aware diarization: Resegmentation using neural end-to-end overlapped speech detection,'' in \emph{IEEE International Conference on Acoustics, Speech and Signal Processing (ICASSP)}, 2020, pp. 7114--7118.

\bibitem{cornell:hal-02908241}
S.~Cornell, M.~Omologo, S.~Squartini, and E.~Vincent, ``{Detecting and counting overlapping speakers in distant speech scenarios},'' in \emph{Proc. Interspeech}, Shanghai, China, Oct. 2020.

\bibitem{CORNELL2022101306}
------, ``Overlapped speech detection and speaker counting using distant microphone arrays,'' \emph{Computer Speech \& Language}, vol.~72, p. 101306, 2022.

\bibitem{zheng2021beamtransformer}
S.~Zheng, S.~Zhang, W.~Huang, Q.~Chen, H.~Suo, M.~Lei, J.~Feng, and Z.~Yan, ``Beamtransformer: Microphone array-based overlapping speech detection,'' \emph{arXiv preprint arXiv:2109.04049}, 2021.

\bibitem{lebourdais2023joint}
M.~Lebourdais, T.~Mariotte, M.~Tahon, A.~Larcher, A.~Laurent, S.~Montresor, S.~Meignier, and J.-H. Thomas, ``Joint speech and overlap detection: a benchmark over multiple audio setup and speech domains,'' \emph{arXiv preprint arXiv:2307.13012}, 2023.

\bibitem{9814838}
S.~Chen, C.~Wang, Z.~Chen, Y.~Wu, S.~Liu, Z.~Chen, J.~Li, N.~Kanda, T.~Yoshioka, X.~Xiao, J.~Wu, L.~Zhou, S.~Ren, Y.~Qian, Y.~Qian, J.~Wu, M.~Zeng, X.~Yu, and F.~Wei, ``Wavlm: Large-scale self-supervised pre-training for full stack speech processing,'' \emph{IEEE Journal of Selected Topics in Signal Processing}, vol.~16, no.~6, pp. 1505--1518, 2022.

\bibitem{10094972}
M.~Kunešová and Z.~Zajíc, ``Multitask detection of speaker changes, overlapping speech and voice activity using {wav2vec} 2.0,'' in \emph{IEEE International Conference on Acoustics, Speech and Signal Processing (ICASSP)}, 2023.

\bibitem{NEURIPS2020_92d1e1eb}
A.~Baevski, Y.~Zhou, A.~Mohamed, and M.~Auli, ``wav2vec 2.0: {A} framework for self-supervised learning of speech representations,'' in \emph{Advances in Neural Information Processing Systems}, vol.~33, 2020, pp. 12\,449--12\,460.

\bibitem{9414677}
Z.-Q. Wang and D.~Wang, ``Count and separate: Incorporating speaker counting for continuous speaker separation,'' in \emph{IEEE International Conference on Acoustics, Speech and Signal Processing (ICASSP)}, 2021, pp. 11--15.

\bibitem{yousefi2021real}
M.~Yousefi and J.~H. Hansen, ``Real-time speaker counting in a cocktail party scenario using attention-guided convolutional neural network,'' \emph{arXiv preprint arXiv:2111.00316}, 2021.

\bibitem{kanda2020joint}
N.~Kanda, Y.~Gaur, X.~Wang, Z.~Meng, Z.~Chen, T.~Zhou, and T.~Yoshioka, ``Joint speaker counting, speech recognition, and speaker identification for overlapped speech of any number of speakers,'' \emph{arXiv preprint arXiv:2006.10930}, 2020.

\bibitem{eliav2024concurrent}
A.~Eliav and S.~Gannot, ``Concurrent speaker detection: A multi-microphone transformer-based approach,'' in \emph{European Signal Processing Conference (EUSIPCO)}, Lyon, France, Aug. 2024.

\bibitem{bredin2021end}
H.~Bredin and A.~Laurent, ``End-to-end speaker segmentation for overlap-aware resegmentation,'' \emph{arXiv preprint arXiv:2104.04045}, 2021.

\bibitem{PORIA201798}
S.~Poria, E.~Cambria, R.~Bajpai, and A.~Hussain, ``A review of affective computing: {F}rom unimodal analysis to multimodal fusion,'' \emph{Information fusion}, vol.~37, pp. 98--125, 2017.

\bibitem{10447462}
S.~Wu, C.~Wang, H.~Chen, Y.~Dai, C.~Zhang, R.~Wang, H.~Lan, J.~Du, C.-H. Lee, J.~Chen, S.~M. Siniscalchi, O.~Scharenborg, Z.-Q. Wang, J.~Pan, and J.~Gao, ``The multimodal information based speech processing {(MISP)} 2023 challenge: {A}udio-visual target speaker extraction,'' in \emph{IEEE International Conference on Acoustics, Speech and Signal Processing (ICASSP)}, 2024, pp. 8351--8355.

\bibitem{10380551}
S.~Cheng, Z.~Ning, J.~Hu, J.~Liu, W.~Yang, L.~Wang, H.~Yu, and W.~Liu, ``G-fusion: Lidar and camera feature fusion on the ground voxel space,'' \emph{IEEE Access}, vol.~12, pp. 4127--4138, 2024.

\bibitem{alameda2024socially}
X.~Alameda-Pineda, A.~Addlesee, D.~H. Garc{\'\i}a, C.~Reinke, S.~Arias, F.~Arrigoni, A.~Auternaud, L.~Blavette, C.~Beyan, L.~G. Camara, \emph{et~al.}, ``Socially pertinent robots in gerontological healthcare,'' \emph{arXiv preprint arXiv:2404.07560}, 2024.

\bibitem{10064301}
M.~Kyoung, H.~Jeon, and K.~Park, ``Audio-visual overlapped speech detection for spontaneous distant speech,'' \emph{IEEE Access}, vol.~11, pp. 27\,426--27\,432, 2023.

\bibitem{mitchell2023study}
D.~A. Mitchell and B.~Rafaely, ``Study of speaker localization under dynamic and reverberant environments,'' \emph{arXiv preprint arXiv:2311.16927}, 2023.

\bibitem{10447683}
C.~Murdock, I.~Ananthabhotla, H.~Lu, and V.~K. Ithapu, ``Self-motion as supervision for egocentric audiovisual localization,'' in \emph{IEEE International Conference on Acoustics, Speech and Signal Processing (ICASSP)}, 2024, pp. 7835--7839.

\bibitem{donley2021easycom}
J.~Donley, V.~Tourbabin, J.-S. Lee, M.~Broyles, H.~Jiang, J.~Shen, M.~Pantic, V.~K. Ithapu, and R.~Mehra, ``Easycom: An augmented reality dataset to support algorithms for easy communication in noisy environments,'' \emph{arXiv preprint arXiv:2107.04174}, 2021.

\bibitem{10184052}
G.~Li, J.~Deng, M.~Geng, Z.~Jin, T.~Wang, S.~Hu, M.~Cui, H.~Meng, and X.~Liu, ``Audio-visual end-to-end multi-channel speech separation, dereverberation and recognition,'' \emph{IEEE/ACM Transactions on Audio, Speech, and Language Processing}, vol.~31, pp. 2707--2723, 2023.

\bibitem{10094836}
Z.~Wang, S.~Wu, H.~Chen, M.-K. He, J.~Du, C.-H. Lee, J.~Chen, S.~Watanabe, S.~Siniscalchi, O.~Scharenborg, D.~Liu, B.~Yin, J.~Pan, J.~Gao, and C.~Liu, ``The multimodal information based speech processing {(MISP)} 2022 challenge: {A}udio-visual diarization and recognition,'' in \emph{IEEE International Conference on Acoustics, Speech and Signal Processing (ICASSP)}, 2023.

\bibitem{hsu2021hubert}
W.-N. Hsu, B.~Bolte, Y.-H.~H. Tsai, K.~Lakhotia, R.~Salakhutdinov, and A.~Mohamed, ``Hubert: Self-supervised speech representation learning by masked prediction of hidden units,'' \emph{IEEE/ACM Transactions on Audio, Speech, and Language Processing}, vol.~29, pp. 3451--3460, 2021.

\bibitem{Tran_2018_CVPR}
D.~Tran, H.~Wang, L.~Torresani, J.~Ray, Y.~LeCun, and M.~Paluri, ``A closer look at spatiotemporal convolutions for action recognition,'' in \emph{Proceedings of the IEEE Conference on Computer Vision and Pattern Recognition (CVPR)}, June 2018.

\bibitem{Jocher_YOLO_by_Ultralytics_2023}
\BIBentryALTinterwordspacing
G.~Jocher, A.~Chaurasia, and J.~Qiu, ``{YOLO} by {U}ltralytics,'' Jan. 2023. [Online]. Available: \url{https://github.com/ultralytics/ultralytics}
\BIBentrySTDinterwordspacing

\bibitem{9222960}
A.~Gillioz, J.~Casas, E.~Mugellini, and O.~A. Khaled, ``Overview of the transformer-based models for {NLP} tasks,'' in \emph{15th Conference on Computer Science and Information Systems (FedCSIS)}, 2020, pp. 179--183.

\bibitem{vaswani2017attention}
A.~Vaswani, N.~Shazeer, N.~Parmar, J.~Uszkoreit, L.~Jones, A.~N. Gomez, {\L}.~Kaiser, and I.~Polosukhin, ``Attention is all you need,'' \emph{Advances in neural information processing systems (NeurIPS)}, vol.~30, 2017.

\bibitem{gong21b_interspeech}
Y.~Gong, Y.-A. Chung, and J.~Glass, ``{AST: Audio Spectrogram Transformer},'' in \emph{Proc. Interspeech}, 2021, pp. 571--575.

\bibitem{dosovitskiy2020image}
A.~Dosovitskiy, L.~Beyer, A.~Kolesnikov, D.~Weissenborn, X.~Zhai, T.~Unterthiner, M.~Dehghani, M.~Minderer, G.~Heigold, S.~Gelly, \emph{et~al.}, ``An image is worth 16x16 words: Transformers for image recognition at scale,'' in \emph{International Conference on Learning Representations (ICLR)}, 2021.

\bibitem{muller2019does}
R.~M{\"u}ller, S.~Kornblith, and G.~E. Hinton, ``When does label smoothing help?'' \emph{Advances in neural information processing systems (NeurIPS)}, vol.~32, 2019.

\bibitem{galdran2020cost}
A.~Galdran, J.~Dolz, H.~Chakor, H.~Lombaert, and I.~Ben~Ayed, ``Cost-sensitive regularization for diabetic retinopathy grading from eye fundus images,'' in \emph{Medical Image Computing and Computer Assisted Intervention (MICCAI)}, 2020, pp. 665--674.

\bibitem{Lin_2017_ICCV}
T.-Y. Lin, P.~Goyal, R.~Girshick, K.~He, and P.~Dollar, ``Focal loss for dense object detection,'' in \emph{Proceedings of the IEEE International Conference on Computer Vision (ICCV)}, Oct. 2017.

\bibitem{10.1007/11677482_3}
J.~Carletta, S.~Ashby, S.~Bourban, M.~Flynn, M.~Guillemot, T.~Hain, J.~Kadlec, V.~Karaiskos, W.~Kraaij, M.~Kronenthal, G.~Lathoud, M.~Lincoln, A.~Lisowska, I.~McCowan, W.~Post, D.~Reidsma, and P.~Wellner, \emph{Machine Learning for Multimodal Interaction}.\hskip 1em plus 0.5em minus 0.4em\relax Springer Berlin Heidelberg, 2006, ch. The {AMI} Meeting Corpus: A Pre-announcement, pp. 28--39.

\end{thebibliography}

\end{document}